\documentclass{article}
\usepackage[dvips]{graphicx}
\begin{document}

\title{ 
Translation Symmetry Breaking in the One-Component Plasma on the Cylinder
}

\author{
L. {\v S}amaj$^1$, J. Wagner$^1$, and P. Kalinay$^{1,2}$
}

\maketitle

\begin{abstract}
The two-dimensional one-component plasma, i.e. the system of pointlike 
charged particles embedded in a homogeneous neutralizing background, 
is studied on the surface of a cylinder of finite circumference, 
or equivalently in a semiperiodic strip of finite width.
The model has been solved exactly by Choquard et al. at the
free-fermion coupling $\Gamma=2$: in the thermodynamic limit
of an infinitely long strip, the particle density turns out to be 
a nonconstant periodic function in space and the system exhibits 
long-range order of the Wigner-crystal type.
The aim of this paper is to describe, qualitatively as well as
quantitatively, the crystalline state for a larger set of couplings 
$\Gamma=2 \gamma$ ($\gamma=1,2\ldots$ a positive integer)
when the plasma is mappable onto a one-dimensional fermionic theory.
The fermionic formalism, supplemented by some periodicity assumptions,
reveals that the density profile results from a hierarchy of Gaussians 
with a uniform variance but with different amplitudes.
The number and spatial positions of these Gaussians within an elementary
cell depend on the particular value of $\gamma$.
Analytic results are supported by the exact solution at $\gamma=1$ 
($\Gamma=2$) and by exact finite-size calculations at $\gamma=2,3$.
\end{abstract}

\medskip


\vfill

\noindent $^1$ 
Institute of Physics, Slovak Academy of Sciences,
D\'ubravsk\'a cesta 9, 845 11 Bratislava, Slovak Republic

\noindent $^2$
Courant Institute of Mathematical Sciences, New York University,
New York, NY 10012

\newpage

\renewcommand{\theequation}{1.\arabic{equation}}
\setcounter{equation}{0}

\section{INTRODUCTION}
According to the laws of electrostatics, the Coulomb potential $v$ 
at a spatial position ${\bf r}\in R^{\nu}$ of the $\nu$-dimensional 
Euclidean space, induced by a unit charge at the origin ${\bf 0}$, 
is defined as the solution of the Poisson equation
\begin{equation} \label{1.1}
\Delta v({\bf r}) = - s_{\nu} \delta({\bf r})
\end{equation}
where $s_{\nu}$ is the surface area of the unit sphere in $R^{\nu}$.
The pair interaction energy of particles with charges 
$q$ and $q'$, localized at the respective positions ${\bf r}$ 
and ${\bf r}'$, is given by
\begin{equation} \label{1.2}
v({\bf r},q\vert {\bf r}',q') = q q' v(\vert {\bf r}-{\bf r}'\vert)
\end{equation}
In one dimension (1D), $s_1=2$ and the solution of (\ref{1.1}) reads
\begin{equation} \label{1.3}
v(x) = - \vert x \vert , \quad \quad \nu=1
\end{equation}
In 2D, $s_2=2\pi$ and the solution of (\ref{1.1}), subject to the
boundary condition $\nabla v({\bf r})\to {\bf 0}$ as 
$\vert {\bf r} \vert \to \infty$, reads
\begin{equation} \label{1.4}
v({\bf r}) = - \ln \left( \frac{\vert {\bf r}\vert}{r_0} \right),
\quad \quad \nu=2
\end{equation}
where $r_0$ is a free length constant which fixes the zero
point of the potential.
The Coulomb potential defined by Eq. (\ref{1.1}) exhibits in the Fourier 
${\bf k}$-space the characteristic singular $\vert {\bf k}\vert^{-2}$ form.
This maintains many generic properties, like the sum rules \cite{Martin},
of ``real'' 3D Coulomb systems with the interaction potential
$v({\bf r})=1/\vert {\bf r}\vert$, ${\bf r}\in R^3$.

The present paper deals with the equilibrium properties of the classical 
(i.e. non-quantum) one-component plasma, sometimes called jellium,
formulated in 1D or quasi-1D domains.
The jellium model consists of only one mobile pointlike particle species 
of charge $q$ embedded in a fixed background of charge $-q$ and density $n$
such that the system as a whole is neutral.

Thermodynamics of the 1D jellium has been obtained exactly
a long time ago by Baxter \cite{Baxter}.
It was proven subsequently that the 1D jellium is never in a fluid state, 
but forms a Wigner crystal \cite{Kunz,Brascamp}.
In particular, choosing the free (hard walls) boundary conditions 
and going to the infinite volume limit, the one-particle density becomes 
periodic in space with period $1/n$.
This long-range order is present for all densities $n$ and all temperatures.
Although the 1D jellium is not in a fluid state, it behaves as a conductor
in the sense that arbitrary boundary charges are perfectly screened by means 
of a global transport of the particle lattice in the background, 
with no additional polarization in the bulk \cite{Lugrin}.

Translation symmetry breaking was documented also on a quasi-1D system,
namely the 2D one-component plasma living on the surface of a cylinder of
circumference $W$ \cite{Choquard1}.
This system is exactly solvable at the dimensionless coupling
constant $\Gamma=2$ \cite{Choquard2}.
In the thermodynamic limit of an infinitely long cylinder, the one-particle 
density is given by an array of equidistant identical Gaussians along
the cylinder's axis, with period $1/(nW)$.

In 1D and quasi-1D Coulomb systems, the variance of the charge 
in an interval $I$ remains uniformly bounded as 
$\vert I\vert \to \infty$.
The existence of periodic structures is related to this boundedness 
of the charge fluctuations \cite{Aizenman,Jancovici1}.

The present work proceeds in the study of the 2D jellium on
the cylinder surface \cite{Choquard1,Choquard2}.
Our aim is to describe, qualitatively as well as quantitatively,
the crystalline state for a larger set of couplings
$\Gamma=2\gamma$ ($\gamma=1,2,\ldots$ a positive integer).
At these couplings, the underlying model is shown to be mappable
onto a 1D anticommuting-field theory following the method of Ref. 
\cite{Samaj1}, and its density profile is expressible in terms 
of the corresponding field correlators.
The assumption of the periodicity of the particle density 
in the thermodynamic limit reveals uniquely that the density profile 
results from a superposition of a hierarchy of nonidentical Gaussians 
with a uniform variance but with different amplitudes.
The number and spatial positions of these Gaussians within an
elementary cell depend on the particular value of $\gamma$.
The analytic results for the crystalline state are supported 
by the exact solution at $\gamma=1$ ($\Gamma=2$) and 
by the exact finite-size calculations at $\gamma=2,3$. 

The paper is organized as follows.
In Section 2, we present basic formulas for the one-component plasma 
living on the cylinder surface.
Section 3 deals with the 1D fermionic representation of the model
for the special values of the coupling constant $\Gamma=2\gamma$ 
($\gamma$ a positive integer). 
Section 4 is devoted to a general analysis of the density profile
in the thermodynamic limit; the Gaussian structure of the crystalline
state is revealed for any value of $\gamma$.
The analytic results are verified in Section 5 on the exact solution 
of the model at $\gamma=1$, and by the exact finite-size 
calculations at $\gamma=2$ and $\gamma=3$.  

\renewcommand{\theequation}{2.\arabic{equation}}
\setcounter{equation}{0}

\section{THE MODEL}
First we define the 2D one-component plasma confined to the surface 
of a cylinder of circumference $W$ and finite length $L$,
in the canonical ensemble.
The cylinder surface can be represented as a 2D semiperiodic rectangle 
domain $\Lambda$ with ${\bf r}=(x,y) \in \Lambda$ if $-L/2\le x\le L/2$ 
(free or hard walls boundary conditions at $x=\pm L/2$) and 
$-W/2\le y\le W/2$ (periodic boundary conditions at $y=\pm W/2$).
It is sometimes useful to use the complex coordinates
$z=x+{\rm i}y$ and ${\bar z}=x-{\rm i}y$.
There are $N$ mobile pointlike particles of charge $q$ in $\Lambda$, 
embedded in a homogeneous background of charge density $\rho_b=-q n$ with
\begin{equation} \label{2.1}
n = \frac{N}{L W}
\end{equation} 
so that the system as a whole is neutral.
The interaction potential between two unit charges at ${\bf r}_1$
and ${\bf r}_2$ is given by the 2D Poisson equation (\ref{1.1})
with the requirement of periodicity along the $y$-axis with period $W$.
Writing the potential as a Fourier series in $y$, one gets \cite{Choquard1}
\begin{equation} \label{2.2}
v({\bf r}_1,{\bf r}_2) = - \ln 
\left\vert 2\, {\rm sinh} \frac{\pi(z_1-z_2)}{W} \right\vert
\end{equation} 
At small distances $\vert {\bf r}_1-{\bf r}_2 \vert << W$,
this potential behaves like the 2D Coulomb potential (\ref{1.4})
with the constant $r_0 = W/(2\pi)$.
At large distances along the cylinder $\vert x_1-x_2\vert >> W$,
this potential behaves like the 1D Coulomb potential 
$-(\pi/W) \vert x_1-x_2\vert$.
We shall need to express the absolute value on the rhs of (\ref{2.2})
formally as the product $g(z_1) g(z_2) \vert f(z_1) - f(z_2) \vert$.
This can be done in two ways:
\begin{eqnarray}
\left\vert 2\, {\rm sinh} \frac{\pi(z_1-z_2)}{W} \right\vert
& = & {\rm e}^{-\frac{\pi}{W}(x_1+x_2)} 
\left\vert {\rm e}^{\frac{2\pi}{W}z_1}-{\rm e}^{\frac{2\pi}{W}z_2}
\right\vert \label{2.3} \\
& = & {\rm e}^{\frac{\pi}{W}(x_1+x_2)} 
\left\vert {\rm e}^{-\frac{2\pi}{W}z_1}-{\rm e}^{-\frac{2\pi}{W}z_2}
\right\vert \label{2.4}
\end{eqnarray} 
The final results cannot depend on the particular choice,
and we shall adopt the representation (\ref{2.3}).
For a given configuration $\{ {\bf r}_1, \ldots, {\bf r}_N \}$
of charges, the total energy of the particle-background system
is given by \cite{Choquard1}
\begin{equation} \label{2.5}
E_N(\{ {\bf r}\}) = q^2 \sum_{j<k} v({\bf r}_j,{\bf r}_k)
+ \pi n q^2 \sum_j x_j^2 + B_N
\end{equation}
where $B_N$ is the background-background interaction constant.

The partition function at inverse temperature $\beta$ is defined by
\begin{equation} \label{2.6}
Z_N = \frac{1}{N!} \int_{\Lambda} \prod_{j=1}^N {\rm d}^2 r_j
{\rm e}^{-\beta E_N(\{ {\bf r}\})}
\end{equation}
It depends on the dimensionless combination $\Gamma = \beta q^2$
called the coupling.
The multiplication of $Z_N$ by a constant does not effect the
particle distribution functions, so for notational convenience
we omit the interaction constant $B_N$ in (\ref{2.5}) and multiply each
volume element ${\rm d}^2 r_j$ in (\ref{2.6}) by $W^{-2}$, to get
\begin{equation} \label{2.7}
Z_N = \frac{1}{N!} \int_{\Lambda} \prod_{j=1}^N
\left[ {\rm d}^2 z_j w(z_j,{\bar z}_j) \right]
\prod_{j<k} \left\vert {\rm e}^{\frac{2\pi}{W}z_j} -
{\rm e}^{\frac{2\pi}{W}z_k} \right\vert^{\Gamma} 
\end{equation}
Here, $w$ is the one-body Boltzmann factor
\begin{equation} \label{2.8}
w(z,{\bar z}) \equiv w(x) = \frac{1}{W^2} \exp \left[ -\pi\Gamma n x^2
-\frac{\pi\Gamma}{W}(N-1)x \right]
\end{equation}
The particle density at point ${\bf r}\in \Lambda$ is defined as
\begin{equation} \label{2.9}
n({\bf r}) = \left\langle \sum_{j=1}^N 
\delta({\bf r}-{\bf r}_j) \right\rangle 
\end{equation}
where $\langle \cdots \rangle$ denotes the usual canonical average.
It can be obtained in a standard way as the functional derivative
\begin{equation} \label{2.10}
n(z,{\bar z}) = w(z,{\bar z}) \frac{\delta}{\delta w(z,{\bar z})} \ln Z_N
\end{equation}
Due to the cylinder geometry of the system, the particle density depends 
only on the $x$-coordinate, $n({\bf r}) \equiv n(x)$, and exhibits 
the reflection symmetry $n(x)=n(-x)$ with respect to the origin $x=0$.
The charge neutrality of the system is equivalent to the condition
\begin{equation} \label{2.11}
\int_{-L/2}^{L/2} {\rm d} x \left[ n(x)-n \right] = 0
\end{equation}

Our task is to determine the particle density profile in the
thermodynamic limit $N,L\to \infty$ (the circumference $W$ of
the cylinder is finite), where the background density $n$ 
given by (\ref{2.1}) stays constant. 
The exact 1D solution of the jellium at any temperature \cite{Kunz}
and the exact 2D solution at the coupling $\Gamma=2$ \cite{Choquard2}
indicate two characteristic features of this density profile:
\begin{itemize}

\item The thermodynamic limit of the density profile depends on
which one of the two subsequences, the particle number
$N=$ even and $N=$ odd integers, is chosen; we denote
by $n^{(e)}(x)$ and $n^{(o)}(x)$ the corresponding density profiles
defined in $-\infty<x<\infty$.
The plots of $n^{(e,o)}(x)$ are expected to be periodic with
a period $\lambda$,
\begin{equation} \label{2.12}
n^{(e,o)}(x\pm \lambda) = n^{(e,o)}(x)
\end{equation} 
The two density profiles are supposed to be the two realizations of 
the same periodic function shifted to one another by a half period,
\begin{equation} \label{2.13}
n^{(e)}(x\pm \lambda/2) = n^{(o)}(x)
\end{equation}

\item The period $\lambda$ is equal to $1/n$ in 1D, independently of 
the temperature.
The exact 2D solution for the coupling $\Gamma=2$ gives $\lambda=1/(nW)$. 
In both cases the elementary cell with the size of the period in
the $x$-direction contains just one particle\footnote{we are grateful 
to B. Jancovici for noticing to us this important fact}.
This motivates us to suggest that the period
\begin{equation} \label{2.14} 
\lambda = \frac{1}{n W}
\end{equation}  
is present for an arbitrary coupling $\Gamma$.
\end{itemize}

\noindent These two working hypothesis will be incorporated into 
an analytic treatment of the model, and subsequently justified 
numerically with a high precision by finite-size calculations. 

\renewcommand{\theequation}{3.\arabic{equation}}
\setcounter{equation}{0}

\section{FERMIONIC REPRESENTATION}
At $\Gamma=2\gamma$ ($\gamma=1,2,\ldots$ a positive integer),
the 2D jellium with the interaction Boltzmann factor
$\prod_{j<k}\vert z_j-z_k\vert^{\Gamma}$ is mappable onto a 
discrete 1D fermionic theory \cite{Samaj1}.
The mapping can be readily extended to the present model.
The partition function (\ref{2.7}) is expressed as an integral
over two sets of Grassman variables 
$\{ \xi_j^{(\alpha)}, \psi_j^{(\alpha)} \}$, each with $\gamma$
components ($\alpha=1,\ldots,\gamma$) defined on a discrete chain
of $N$ sites $j=0,1,\ldots,N-1$ as follows
\begin{eqnarray}
Z_N & = & \int {\cal D}\psi {\cal D}\xi\, {\rm e}^{S(\xi,\psi)}
\label{3.1} \\
S(\xi,\psi) & = & \sum_{j,k=0}^{\gamma(N-1)} \Xi_j w_{jk} \Psi_k
\label{3.2}
\end{eqnarray}
Here ${\cal D}\psi {\cal D}\xi = \prod_{j=0}^{N-1} 
{\rm d}\psi_j^{(\gamma)} \ldots {\rm d}\psi_j^{(1)}
{\rm d}\xi_j^{(\gamma)} \ldots {\rm d}\xi_j^{(1)}$
and the action $S$ involves pair interactions of ``composite''
variables
\begin{equation} \label{3.3}
\Xi_j = \sum_{j_1,\ldots,j_{\gamma}=0\atop (j_1+\ldots+j_{\gamma}=j)}^{N-1}
\xi_{j_1}^{(1)} \ldots \xi_{j_{\gamma}}^{(\gamma)}, \quad \quad
\Psi_k = \sum_{k_1,\ldots,k_{\gamma}=0\atop (k_1+\ldots+k_{\gamma}=k)}^{N-1}
\psi_{k_1}^{(1)} \ldots \psi_{k_{\gamma}}^{(\gamma)}
\end{equation} 
The interaction strengths $w_{jk}$ $[j,k=0,1,\ldots,\gamma(N-1)]$
are given by
\begin{equation} \label{3.4}
w_{jk} = \int_{\Lambda} {\rm d}^2 z\, w(z,{\bar z}) 
\exp\left( \frac{2\pi}{W} j z \right) 
\exp\left( \frac{2\pi}{W} k {\bar z} \right)
\end{equation}
Using the notation 
$\langle \cdots \rangle = \int {\cal D}\psi {\cal D}\xi {\rm e}^S \cdots/Z_N$
for averaging over the anticommuting variables, the particle density
(\ref{2.10}) is expressible in the fermionic form as follows
\begin{equation} \label{3.5}
n(z,{\bar z}) = w(z,{\bar z}) \sum_{j,k=0}^{\gamma(N-1)}
\langle \Xi_j \Psi_k \rangle \exp\left( \frac{2\pi}{W} j z \right) 
\exp\left( \frac{2\pi}{W} k {\bar z} \right)
\end{equation}
The fermionic correlators $\{ \langle \Xi_j \Psi_k \rangle \}$
can be obtained from the partition function (\ref{3.1}) using
relation (\ref{3.2}) via the derivatives
\begin{equation} \label{3.6}
\langle \Xi_j \Psi_k \rangle =
\frac{\partial}{\partial w_{jk}} \ln Z_N
\end{equation}

Inserting the one-body Botzmann factor $w$ of interest (\ref{2.8})
into (\ref{3.4}), and using the orthogonality relation
\begin{equation} \label{3.7}
\int_{-W/2}^{W/2} {\rm d}y\, 
\exp\left\{ \frac{2\pi}{W} {\rm i} (j-k) y \right\}
= W \delta_{jk}
\end{equation}
leads to the diagonalization of the interaction matrix
\begin{equation} \label{3.8}
w_{jk} = w_j \delta_{jk}, \quad \quad
w_j = \frac{1}{W} \int_{-L/2}^{L/2} {\rm d}x\,
\exp\left\{ -2\pi\gamma n x^2 + \frac{2\pi}{W}
\left[ 2 j - \gamma(N-1) \right] x \right\}
\end{equation}
Notice that the diagonal interaction strengths $\{ w_j \}$
possess the symmetry
\begin{equation} \label{3.9}
w_j = w_{\gamma(N-1)-j} \quad \quad
\mbox{for all $j=0,1,\ldots,\gamma(N-1)$}
\end{equation}
With $w_{jk}$ of the form (\ref{3.8}), the action (\ref{3.2})
of the partition function (\ref{3.1}) becomes ``diagonalized'',
\begin{equation} \label{3.10}
Z_N = \int {\cal D}\psi {\cal D}\xi
\exp \left( \sum_{j=0}^{\gamma(N-1)} \Xi_j w_j \Psi_j \right)
\end{equation}
and the fermionic correlators $\{ \langle \Xi_j \Psi_k \rangle \}$
are given by
\begin{equation} \label{3.11}
\langle \Xi_j \Psi_k \rangle = \langle \Xi_j \Psi_j \rangle \delta_{jk},
\quad \quad 
\langle \Xi_j \Psi_j \rangle = \frac{\partial}{\partial w_j} \ln Z_N
\end{equation}
The particle density (\ref{3.5}) takes the form
\begin{equation} \label{3.12}
n(x) = \frac{1}{W^2} \sum_{j=0}^{\gamma(N-1)} \langle \Xi_j \Psi_j \rangle
\exp\left\{ -2\pi\gamma n x^2 + \frac{2\pi}{W}
\left[ 2 j - \gamma(N-1) \right] x \right\}
\end{equation}
where $-\lambda N/2\le x\le \lambda N/2$, $\lambda$ being defined by
Eqs. (\ref{2.1}) and (\ref{2.14}).
The symmetry (\ref{3.9}) of the interaction strengths $\{ w_j \}$
implies an analogous symmetry for the fermionic correlators
\begin{equation} \label{3.13}
\langle \Xi_j \Psi_j \rangle = \langle \Xi_{\gamma(N-1)-j} 
\Psi_{\gamma(N-1)-j} \rangle \quad \quad
\mbox{for all $j=0,1,\ldots,\gamma(N-1)$}
\end{equation}
This relation ensures the mentioned reflection property
of the particle density $n(x)=n(-x)$.

\renewcommand{\theequation}{4.\arabic{equation}}
\setcounter{equation}{0}

\section{GENERAL ANALYSIS}
The explicit formula (\ref{3.12}) for the density profile contains
the unknown set of fermionic correlators 
$\{ \langle \Xi_j \Psi_j \rangle \}$.
In this section we present an analytic treatment of the thermodynamic
$N,L\to \infty$ limit of this formula, supplemented by the 
periodicity assumptions (\ref{2.12}) and (\ref{2.13}) for the
density functions with the period $\lambda$ defined by Eq. (\ref{2.14}).
It will be shown that for any $\gamma$ the periodicity assumptions
determine uniquely the general Gaussian structure of the density
profile, without having at one's disposal the particular values
of the fermionic correlators $\{ \langle \Xi_j \Psi_j \rangle \}$.
The treatment depends technically on whether $\gamma$ is an even 
or odd integer.

\subsection{$\gamma=$ even integer}
For $\gamma$ even, we define the auxiliary integer $M$ as follows
\begin{equation} \label{4.1}
\gamma(N-1) = 2 M
\end{equation}
Let us shift the $j$-index enumeration in Eq. (\ref{3.12}) by $M$,
\begin{equation} \label{4.2}
j=M+l, \quad \quad l=-M,-M+1,\ldots,M
\end{equation}
Now, rescaling appropriately the integration $x$-variable in Eq. (\ref{3.8}), 
the interaction strengths can be written as
\begin{equation} \label{4.3}
w_{M+l} = 
\frac{\exp\left( \frac{2\pi l^2}{\gamma\mu}\right)}{\sqrt{2\gamma\mu}}
\frac{1}{\sqrt{\pi}} 
\int_{-\sqrt{\frac{\pi\gamma}{2\mu}}N}^{\sqrt{\frac{\pi\gamma}{2\mu}}N}
{\rm d}x \, \exp\left\{ -\left( x- \sqrt{\frac{2\pi}{\gamma\mu}}l\right)^2
\right\}
\end{equation}
Here, we have introduced the dimensionless parameter
\begin{equation} \label{4.4}
\mu = n W^2
\end{equation}
which measures the number of particles in a square of side $W$;
the limits $\mu\to 0$ and $\mu\to\infty$ correspond to the extreme
1D (at zero temperature) and 2D versions of the model, respectively.
The symmetry (\ref{3.9}) is equivalent to
\begin{equation} \label{4.5}
w_{M+l} = w_{M-l} \quad \quad
\mbox{for all $l=-M,\ldots,M$}
\end{equation}
which can be verified directly from the explicit representation (\ref{4.3}).
For $l$ finite and in the limit $N\to\infty$
\begin{equation} \label{4.6}
w_{M+l} \sim \frac{1}{\sqrt{2\gamma\mu}}
\exp\left( \frac{2\pi l^2}{\gamma\mu} \right)
\end{equation}
Under the index shift (\ref{4.2}), the density profile (\ref{3.12})
can be expressed as
\begin{equation} \label{4.7}
\frac{n(x)}{n} = \sqrt{\frac{2}{\gamma\mu}} \sum_{l=-M}^M
c_l(M) \exp\left\{ - \frac{2\pi\gamma}{\mu} \left( \frac{x}{\lambda}
-\frac{l}{\gamma} \right)^2 \right\}
\end{equation}
where we have introduced the coefficients
\begin{equation} \label{4.8}
c_l(M) = \langle \Xi_{M+l} \Psi_{M+l} \rangle 
\sqrt{\frac{\gamma}{2\mu}} \exp\left( \frac{2\pi l^2}{\gamma\mu} \right)
\end{equation}
The symmetry of the fermionic correlators (\ref{3.13}) is equivalent to
$\langle \Xi_{M+l} \Psi_{M+l} \rangle = \langle \Xi_{M-l} \Psi_{M-l} \rangle$.
The consequent relation
\begin{equation} \label{4.9}
c_l(M) = c_{-l}(M) \quad \quad 
\mbox{for all $l=-M,\ldots,M$}
\end{equation}
ensures the reflection symmetry $n(x)=n(-x)$.
We see that, even for a finite-size system, the particle density 
(\ref{4.7}) results as a superposition of Gaussians, 
localized equidistantly at positions $x=\lambda l/\gamma$ ($l=-M,\ldots,M$), 
with the uniform variance $\sigma^2=\lambda^2\mu/(4\pi\gamma)$ 
but with different position-dependent amplitudes.
The value-structure of these amplitudes simplifies substantially
in the thermodynamic limit discussed below.

All formal algebra made till now was rigorous. 
To describe the thermodynamic limit of the profile relation (\ref{4.7}),
we adopt the two assumptions presented at the end of Section 2.
In the limit $M\to\infty$, one has to distinguish between the
subsequences of $M$ in (\ref{4.1}) which correspond to $N=$even and
to $N=$odd particle numbers.
Within each of the even and odd $M$-subsequences, the coefficients
$\{ c_l(M) \}$ tend uniformly to their asymptotic values denoted by
$\{ c_l^{(e)} \}$ and $\{ c_l^{(o)} \}$, respectively.
Thence,
\begin{equation} \label{4.10}
\frac{n^{(e,o)}(x)}{n} = \sqrt{\frac{2}{\gamma\mu}} 
\sum_{l=0,\pm 1,\ldots} c^{(e,o)}_l \exp\left\{ - \frac{2\pi\gamma}{\mu} 
\left( \frac{x}{\lambda}-\frac{l}{\gamma} \right)^2 \right\}
\end{equation}
The analogue of the symmetry relation (\ref{4.9}) takes the form
\begin{equation} \label{4.11}
c^{(e,o)}_l = c^{(e,o)}_{-l} \quad \quad \mbox{for all $l=0,\pm 1,\ldots$}
\end{equation}
The periodicity assumption (\ref{2.12}) implies
\begin{equation} \label{4.12}
c_l^{(e,o)} = c_{l\pm \gamma}^{(e,o)}
\quad \quad \mbox{for all $l=0,\pm 1,\ldots$}
\end{equation}
The shift condition (\ref{2.13}) between the ``even'' and ``odd''
states leads to the relations
\begin{equation} \label{4.13}
c_l^{(e)} = c_{l\pm \frac{\gamma}{2}}^{(o)}
\quad \quad \mbox{for all $l=0,\pm 1,\ldots$}
\end{equation}
Based on Eqs. (\ref{4.11})-(\ref{4.13}) we conclude that there exist
$\gamma/2+1$ independent asymptotic amplitudes 
$C_0, C_1,\ldots, C_{\frac{\gamma}{2}}$ such that
\begin{equation} \label{4.14}
c_l^{(e)} = C_l, \quad \quad c_l^{(o)} = C_{\frac{\gamma}{2}-l} \quad \quad
\mbox{for $l=0,1,\ldots,\frac{\gamma}{2}$}
\end{equation}
All other coefficients can be generated from the basic set (\ref{4.14})
by using the symmetry relations (\ref{4.11}) and (\ref{4.12}).
The values of the asymptotic $C$-amplitudes depend on the dimensionless 
parameter $\mu$ of Eq. (\ref{4.4}).
They are constrained by the neutrality condition (\ref{2.11}),
written for $x$ ranging over one period as follows
\begin{equation} \label{4.15}
\int_0^{\lambda} {\rm d} x 
\left[ \frac{n^{(e,o)}(x)}{n} - 1 \right] = 0
\end{equation}
Simple algebra gives
\begin{equation} \label{4.16}
C_0 +2(C_1 + \cdots + C_{\frac{\gamma}{2}-1}) +C_{\frac{\gamma}{2}} = \gamma
\end{equation}

For instance, in the $\gamma=2$ case, there are two amplitudes
$C_0$ and $C_1$ such that
\begin{eqnarray}
c_l^{(e)} & = &  \left\{
\begin{array}{ll}
C_0 & \mbox{for $l$ even} \cr 
C_1 & \mbox{for $l$ odd}
\end{array} \right.
\label{4.17} \\
c_l^{(o)} & = &  \left\{
\begin{array}{ll}
C_1 & \mbox{for $l$ even} \cr 
C_0 & \mbox{for $l$ odd}
\end{array} \right.
\label{4.18}
\end{eqnarray}
The amplitudes are constrained by
\begin{equation} \label{4.19}
C_0(\mu) + C_1(\mu) = 2 
\end{equation}

\subsection{$\gamma=$ odd integer}
For $\gamma$ odd, the fermionic formalism needs to be developed separately
for the thermodynamic-limit subsequence with $N$ even and $N$ odd.

When the particle number $N$ is even, the product
\begin{equation} \label{4.20}
\gamma(N-1) = 2M + 1
\end{equation}
with some integer $M$.
The shift of the $j$-index enumeration in Eq. (\ref{3.12})
\begin{equation} \label{4.21}
j = M + \frac{1}{2} + l, \quad \quad
l = -\left( M+\frac{1}{2} \right), -M+\frac{1}{2}, \ldots, M+\frac{1}{2}
\end{equation}
leads to the interaction strengths
\begin{equation} \label{4.22}
w_{M+\frac{1}{2}+l} = 
\frac{\exp\left( \frac{2\pi l^2}{\gamma\mu}\right)}{\sqrt{2\gamma\mu}}
\frac{1}{\sqrt{\pi}} 
\int_{-\sqrt{\frac{\pi\gamma}{2\mu}}N}^{\sqrt{\frac{\pi\gamma}{2\mu}}N}
{\rm d}x \, \exp\left\{ -\left( x- \sqrt{\frac{2\pi}{\gamma\mu}}l\right)^2
\right\}
\end{equation}
and to the coefficients
\begin{equation} \label{4.23}
c_l(M) = \langle \Xi_{M+\frac{1}{2}+l} \Psi_{M+\frac{1}{2}+l} \rangle 
\sqrt{\frac{\gamma}{2\mu}} \exp\left( \frac{2\pi l^2}{\gamma\mu} \right)
\end{equation}
possessing the symmetry $c_l(M)=c_{-l}(M)$.
In the limit $N\to\infty$ (keeping $N$ even), one arrives at
the density profile of the ``even'' state
\begin{equation} \label{4.24}
\frac{n^{(e)}(x)}{n} = \sqrt{\frac{2}{\gamma\mu}} 
\sum_{l=\pm\frac{1}{2},\pm\frac{3}{2},\ldots} 
c^{(e)}_l \exp\left\{ - \frac{2\pi\gamma}{\mu} 
\left( \frac{x}{\lambda}-\frac{l}{\gamma} \right)^2 \right\}
\end{equation}
The asymptotic coefficients satisfy the symmetry relation
\begin{equation} \label{4.25}
c_l^{(e)} = c_{-l}^{(e)} \quad \quad
\mbox{for all $l=\pm\frac{1}{2}, \pm\frac{3}{2}, \ldots$}
\end{equation}
and the periodicity condition
\begin{equation} \label{4.26}
c_l^{(e)} = c_{l\pm \gamma}^{(e)} \quad \quad
\mbox{for all $l=\pm\frac{1}{2}, \pm\frac{3}{2}, \ldots$}
\end{equation}
implied by Eqs. (\ref{2.12}) and (\ref{2.14}).
As a consequence, there exist $(\gamma+1)/2$ independent amplitudes
$C_{\frac{1}{2}}, C_{\frac{3}{2}},\ldots,C_{\frac{\gamma}{2}}$
such that
\begin{equation} \label{4.27}
c_l^{(e)} = C_l \quad \quad
\mbox{for $l=\frac{1}{2},\frac{3}{2},\ldots,\frac{\gamma}{2}$}
\end{equation}
All other coefficients can be generated from the basic set (\ref{4.27})
with the aid of the symmetry relations (\ref{4.25}) and (\ref{4.26}).

When the particle number $N$ is odd, the product $\gamma(N-1)$
is expressible similarly as in Eq. (\ref{4.1}), so that we can proceed 
along the lines of Section 4.1.
The density profile of the ``odd'' state is again of the form (\ref{4.10})
\begin{equation} \label{4.28}
\frac{n^{(o)}(x)}{n} = \sqrt{\frac{2}{\gamma\mu}} 
\sum_{l=0,\pm 1,\ldots} c^{(o)}_l \exp\left\{ - \frac{2\pi\gamma}{\mu} 
\left( \frac{x}{\lambda}-\frac{l}{\gamma} \right)^2 \right\}
\end{equation}
where the asymptotic coefficients exhibit the symmetries
$c_l^{(o)} = c_{-l}^{(o)}$ and $c_l^{(o)} = c_{l\pm \gamma}^{(o)}$
for all $l=0,\pm 1,\ldots$.
The shift condition (\ref{2.13}), when applied to the representations
(\ref{4.24}) and (\ref{4.28}), gives 
$c_l^{(e)} = c_{l\pm\frac{\gamma}{2}}^{(o)}$.
In view of Eq. (\ref{4.27}), the basic set of the odd coefficients is
given by
\begin{equation} \label{4.29}
c_l^{(o)} = C_{\frac{\gamma}{2}-l} \quad \quad
\mbox{for $l=0,1,\ldots,\frac{\gamma-1}{2}$}
\end{equation}
The neutrality conditions of type (\ref{4.15}) constraint
the $C$-amplitudes as follows
\begin{equation} \label{4.30}
2\left[ C_{\frac{1}{2}}(\mu) + \cdots + C_{\frac{\gamma}{2}-1}(\mu) \right]
+ C_{\frac{\gamma}{2}}(\mu) = \gamma
\end{equation}

For $\gamma=1$, one has the simple result
\begin{eqnarray}
c_l^{(e)} & = & 1 \quad 
\mbox{for all $l=\pm\frac{1}{2},\pm\frac{3}{2},\ldots$} \label{4.31} \\
c_l^{(o)} & = & 1 \quad 
\mbox{for all $l=0,\pm 1,\ldots$} \label{4.32} 
\end{eqnarray}

For $\gamma=3$, the above scheme results in
\begin{eqnarray}
c_l^{(e)} & = & \left\{
\begin{array}{ll}
C_{\frac{1}{2}} & \mbox{for $l=\pm\frac{1}{2}+3k$} \cr
C_{\frac{3}{2}} & \mbox{for $l=\frac{3}{2}+3k$} 
\end{array} \right.
\label{4.33} \\
c_l^{(o)} & = & \left\{
\begin{array}{ll}
C_{\frac{3}{2}} & \mbox{for $l=3k$} \cr
C_{\frac{1}{2}} & \mbox{for $l=\pm 1+3k$} 
\end{array} \right.
\label{4.34}
\end{eqnarray}
where $k$ is an arbitrary integer.
The $\mu$-dependent amplitudes are constrained by
\begin{equation} \label{4.35}
2 C_{\frac{1}{2}}(\mu) + C_{\frac{3}{2}}(\mu) = 3
\end{equation}

\renewcommand{\theequation}{5.\arabic{equation}}
\setcounter{equation}{0}

\section{FINITE-SIZE CALCULATIONS}
In this section, we check the obtained analytic results by
the exact finite-$N$ calculations.
The crucial problem is to determine the dependence of the
partition function $Z_N(\gamma)$, given as the integral over
anticommuting variables by Eq. (\ref{3.10}), on the set
of interaction strengths $\{ w_j \}_{j=0}^{\gamma(N-1)}$.
Having at one's disposal this dependence, the consequent fermionic
correlators [generated by using Eq. (\ref{3.11})] 
determine the $c$-coefficients of interest via relations (\ref{4.8}) 
[($\gamma$ even, $N$ arbitrary) or ($\gamma$ odd, $N$ odd)] 
and (\ref{4.23}) ($\gamma$ odd, $N$ even).

For $\gamma=1$, one has the simple result \cite{Jancovici2}
\begin{eqnarray}
Z_N(1) & = & w_0 w_1 \cdots w_{N-1} \label{5.1} \\
\langle \Xi_j \Psi_j \rangle & = & \frac{1}{w_j}, \quad \quad
j=0,1,\ldots,N-1  \label{5.2}
\end{eqnarray} 
In the thermodynamic $N\to\infty$ limit, after simple algebra
both even (\ref{4.31}) and odd (\ref{4.32}) types of 
the $c$-coefficients are reproduced exactly. 

For larger values of $\gamma$, the partition function is a more
complicated function of interaction strengths whose complexity
increases with increasing the particle number $N$.
The methods for a systematic generation of $Z_N(\gamma)$,
realized in practice through computer languages like {\it Fortran}, 
are summarized and further developed in Ref. \cite{Samaj2}.
With the aid of these methods we were able to go up to $N=12$
particles for $\gamma=2$ and up to $N=9$ particles for $\gamma=3$. 

\subsection{Results for $\gamma=2$}
The finite-size results for $\gamma=2$ are summarized in Figs. 1-3.
Their discussion follows the analysis of Subsection 4.1.
The asymptotic amplitudes $C_0$ and $C_1$ are defined by relations
(\ref{4.17}) and (\ref{4.18}).

Taking the parameter $\mu=3$, the density profiles for the subsequence 
with the particle number $N$ even and the one with $N$ odd are plotted
in Figs. 1a and 1b, respectively.
In the units of the period $\lambda=1$, hard walls are localized
at $x=\pm N/2$ where the boundary effects dominate.
On the other hand, close to the center $x=0$, the particle density
exhibits the characteristic periodic behavior as from relatively
small particle numbers.
The density at $x=0$ has the minimum for $N$ even and the maximum for $N$ odd.
This means that the asymptotic amplitudes fulfill the inequality $C_0<C_1$.
The Gaussians with the smaller asymptotic amplitude $C_0$ are localized at 
the minimum points of the density profile, and so they modify the density 
plot given by the equidistant array of the Gaussians corresponding to 
the larger asymptotic amplitude $C_1$ (and localized at the maximum points 
of the density profile) only marginally, without generating new extreme points.
We observe numerically this phenomenon for any value of $\mu$.

Taking the same value of the parameter $\mu=3$, the plots of the coefficients
$\{ c_0,c_1,c_2,c_3\}$, reflecting the amplitudes of the Gaussians 
close to the center, versus $1/N$ are pictured in Fig. 2
where only the subsequence with $N$ odd is considered.
As the particle number $N$ increases, $c_0\to c_2$ and $c_1\to c_3$
as is expected.
As can be seen the convergence of the $c$-coefficients to their asymptotic
$C_0$ and $C_1$ values turns out to be fast.

The dependence of the asymptotic amplitudes $C_0$ and $C_1$,
constrained by $C_0+C_1=2$, on the parameter $\mu$ is shown in Fig. 3.
The finite-size errors in determining these amplitudes are deducible
directly from the differences $c_0-c_2$ and $c_1-c_3$ at the largest
$N=12$ particle number.
The error bars increase with increasing $\mu$, they are however still
so small that we do not present them in the figure in order to
keep the clarity of the presentation.
In the limit $\mu\to 0$, which corresponds to the 1D jellium
at zero temperature, $C_0=0$ and $C_1=2$; there exists consequently
only one set of Gaussians (more precisely, the $\delta$-function peaks), 
in agreement with the exact 1D solution \cite{Kunz}.
In the limit $\mu\to \infty$, which corresponds to the bulk 2D jellium, 
both asymptotic amplitudes tend to unity as it should be. 

\subsection{Results for $\gamma=3$}
The finite-size results for $\gamma=3$ are summarized in Figs. 4-6.
We do not comment these figures since they provide similar information 
as the previous ones for $\gamma=2$.
We only recall that the asymptotic amplitudes $C_{1/2}$ 
and $C_{3/2}$, constrained by $2 C_{1/2}+C_{3/2}=3$,
are defined by Eqs. (\ref{4.33}) and (\ref{4.34}).
\medskip

One concludes that the finite-size calculations confirm with a high
accuracy the predicted Gaussian structure of the crystalline state.
The structure of the crystalline state for noninteger values of $\gamma$ 
is an open problem.

\section*{ACKNOWLEDGMENTS}
We thank B. Jancovici for stimulating discussions.
The support by a VEGA grant is acknowledged.

\newpage

\begin{figure}[t]
\includegraphics[scale=0.98]{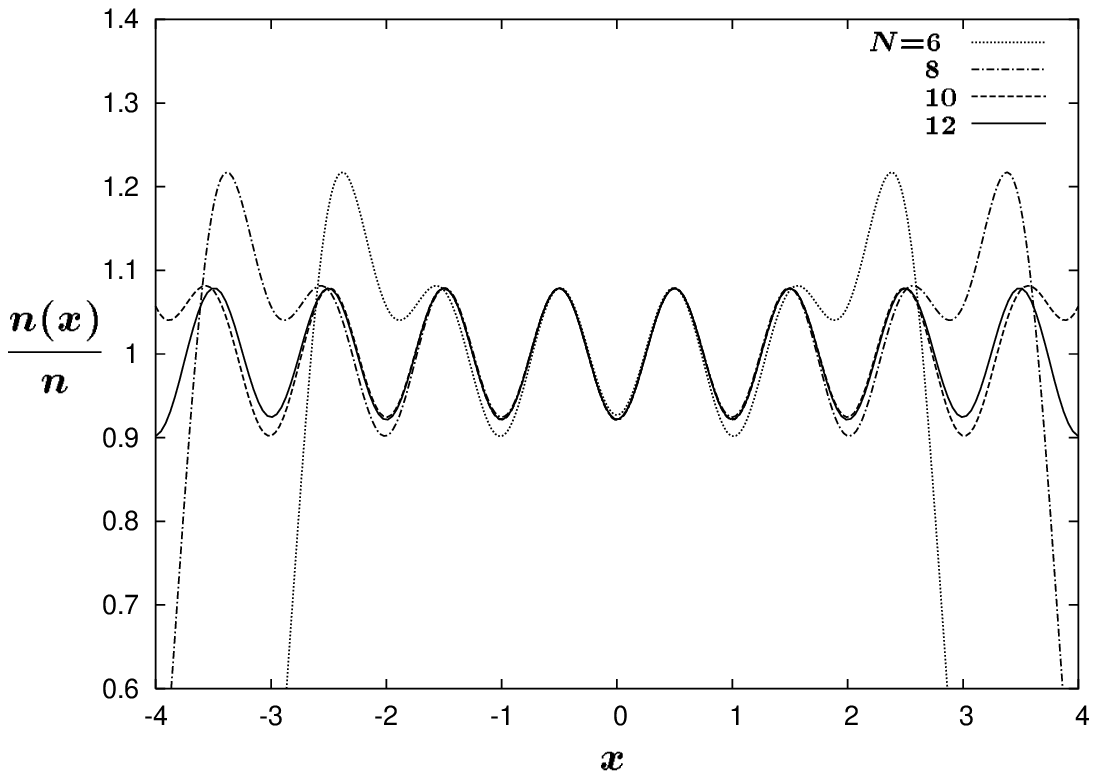}
\begin{center}
\mbox{Fig. 1a. Density profiles for $N$ even: $\gamma=2$, $\mu=3$.}
\end{center}
\end{figure}

\begin{figure}[b]
\includegraphics[scale=0.98]{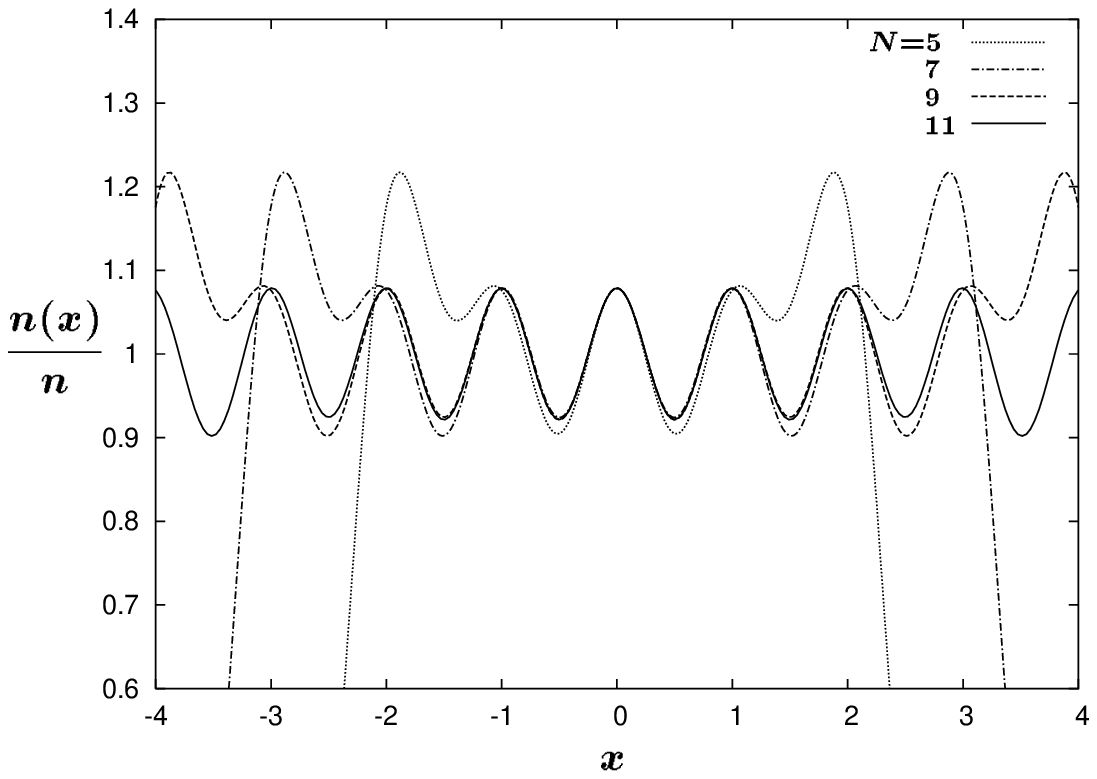}
\begin{center}
\mbox{Fig. 1b. Density profiles for $N$ odd: $\gamma=2$, $\mu=3$.}
\end{center}
\end{figure}

\begin{figure}[t]
\includegraphics[scale=0.96]{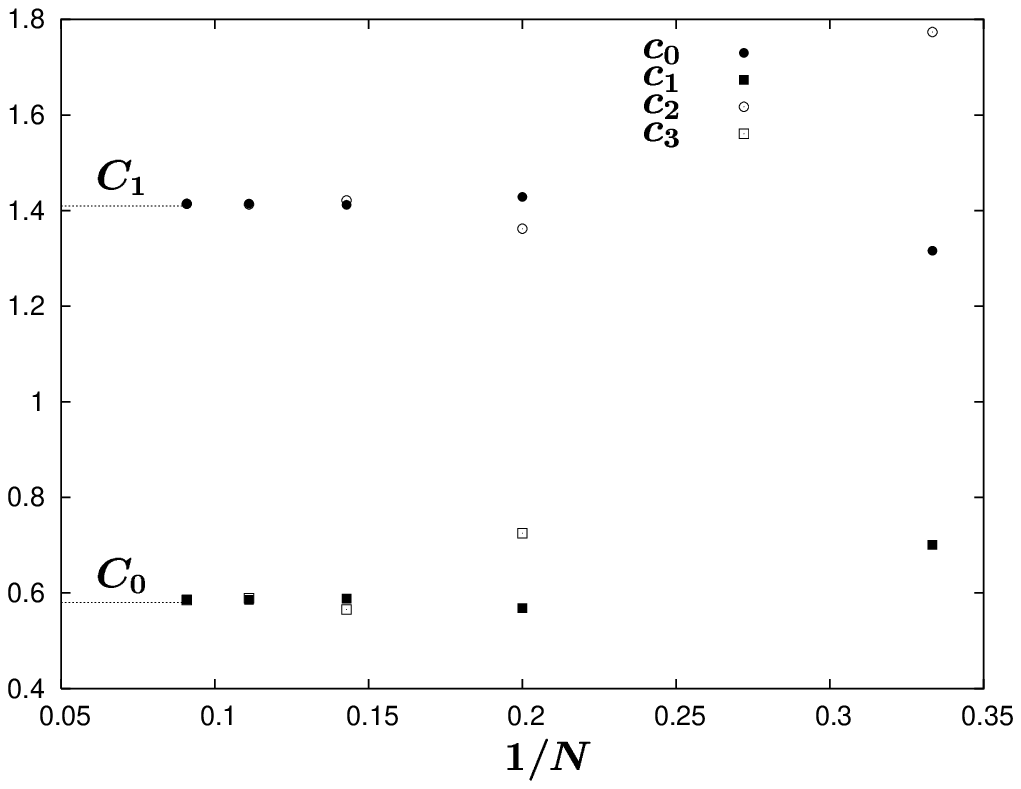}
\begin{center}
\mbox{Fig. 2. The dependence of the $c$-coefficients 
on $1/N$: $\gamma=2$, $\mu=3$.}
\end{center}
\end{figure}

\begin{figure}[b]
\includegraphics[scale=0.96]{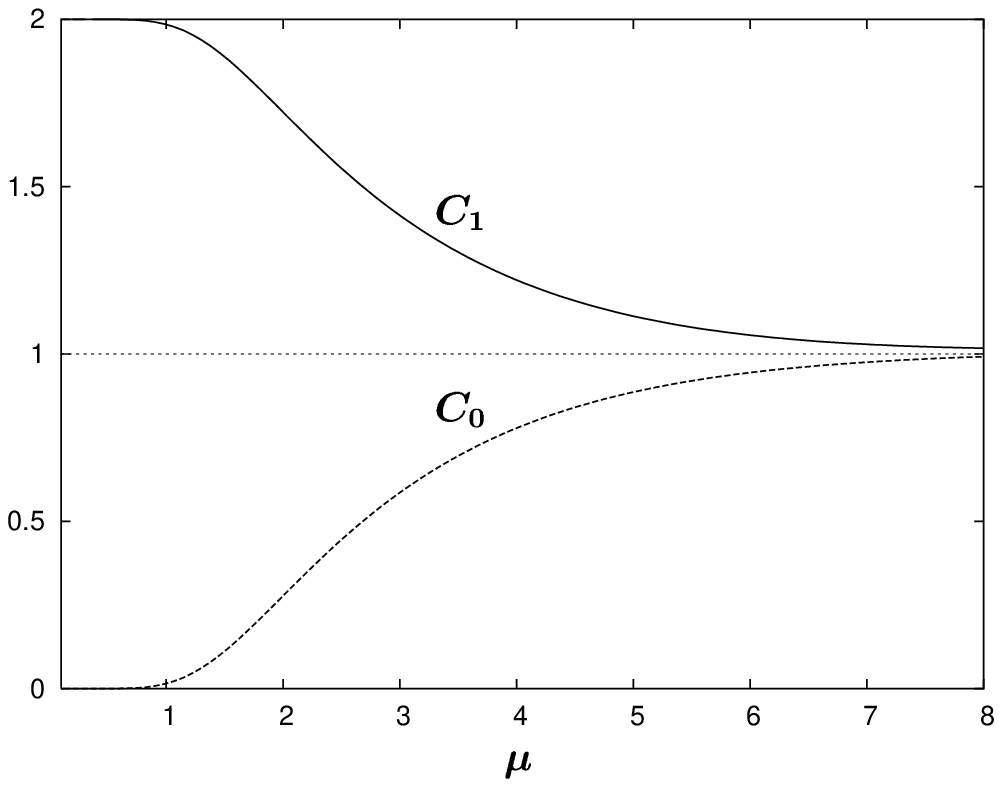}
\begin{center}
\mbox{Fig. 3. The plots of the asymptotic $C$-amplitudes
vs. $\mu$: $\gamma=2$.}
\end{center}
\end{figure}

\begin{figure}[t]
\includegraphics[scale=0.98]{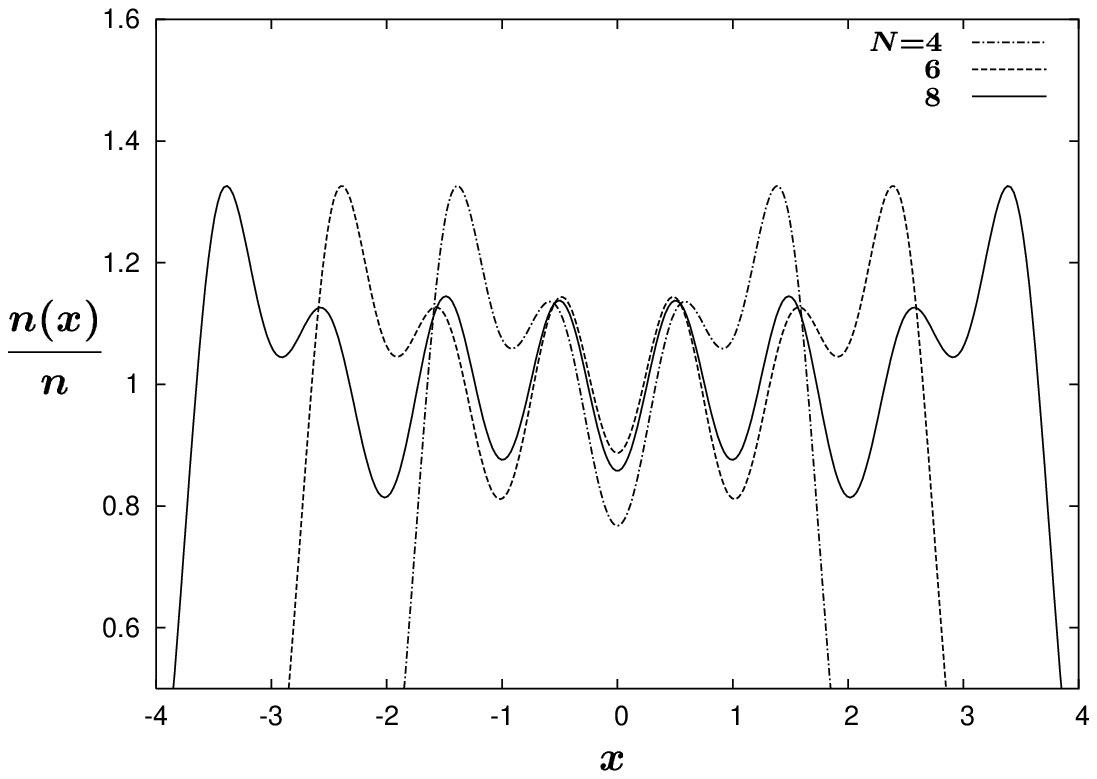}
\begin{center}
\mbox{Fig. 4a. Density profiles for $N$ even: $\gamma=3$, $\mu=3$.}
\end{center}
\end{figure}

\begin{figure}[b]
\includegraphics[scale=0.98]{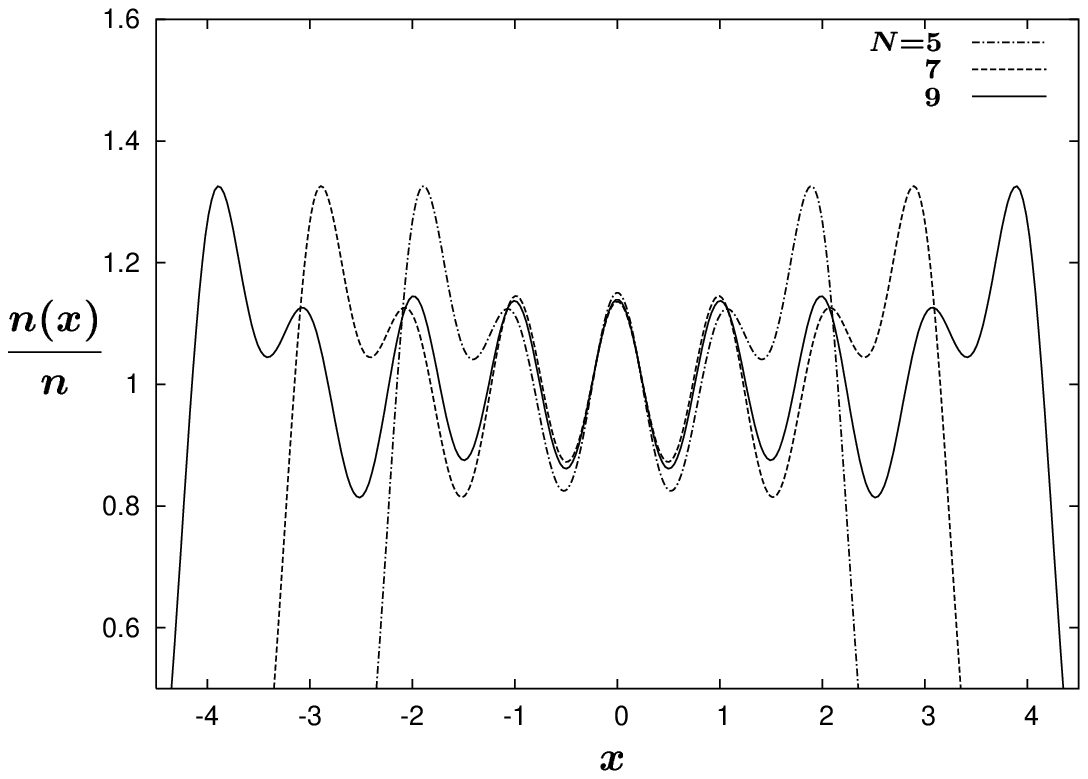}
\begin{center}
\mbox{Fig. 4b. Density profiles for $N$ odd: $\gamma=3$, $\mu=3$.}
\end{center}
\end{figure}

\begin{figure}[t]
\includegraphics[scale=0.96]{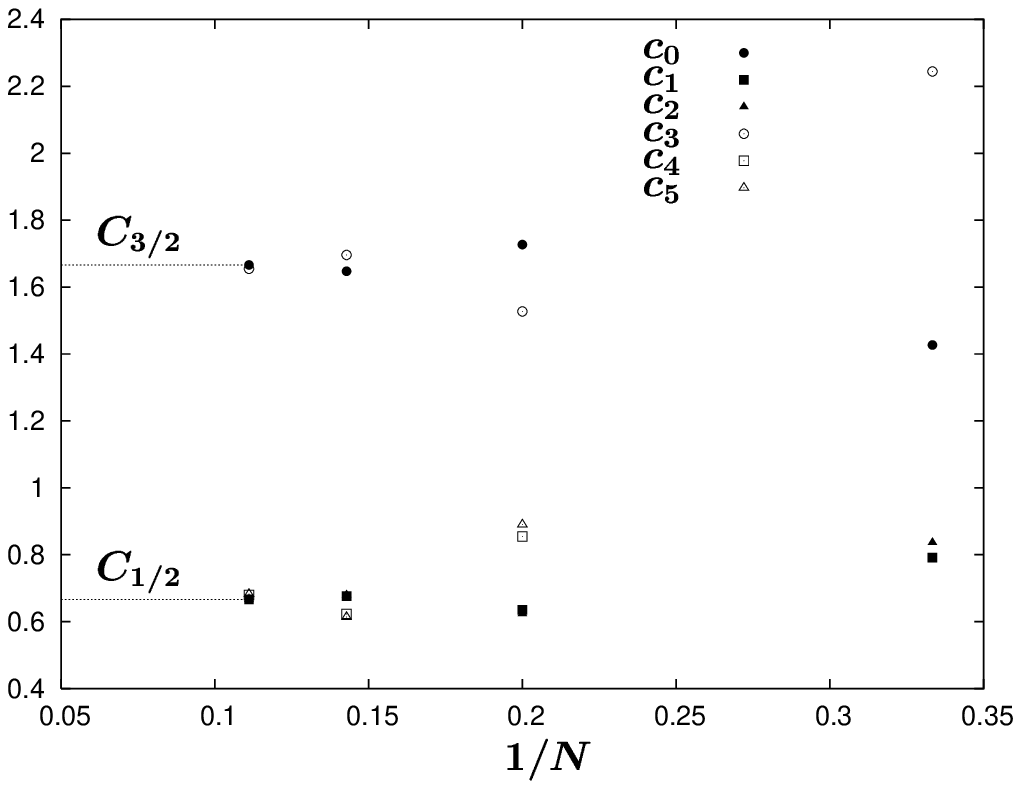}
\begin{center}
\mbox{Fig. 5. The dependence of the $c$-coefficients 
on $1/N$: $\gamma=3$, $\mu=3$.}
\end{center}
\end{figure}

\begin{figure}[b]
\includegraphics[scale=0.96]{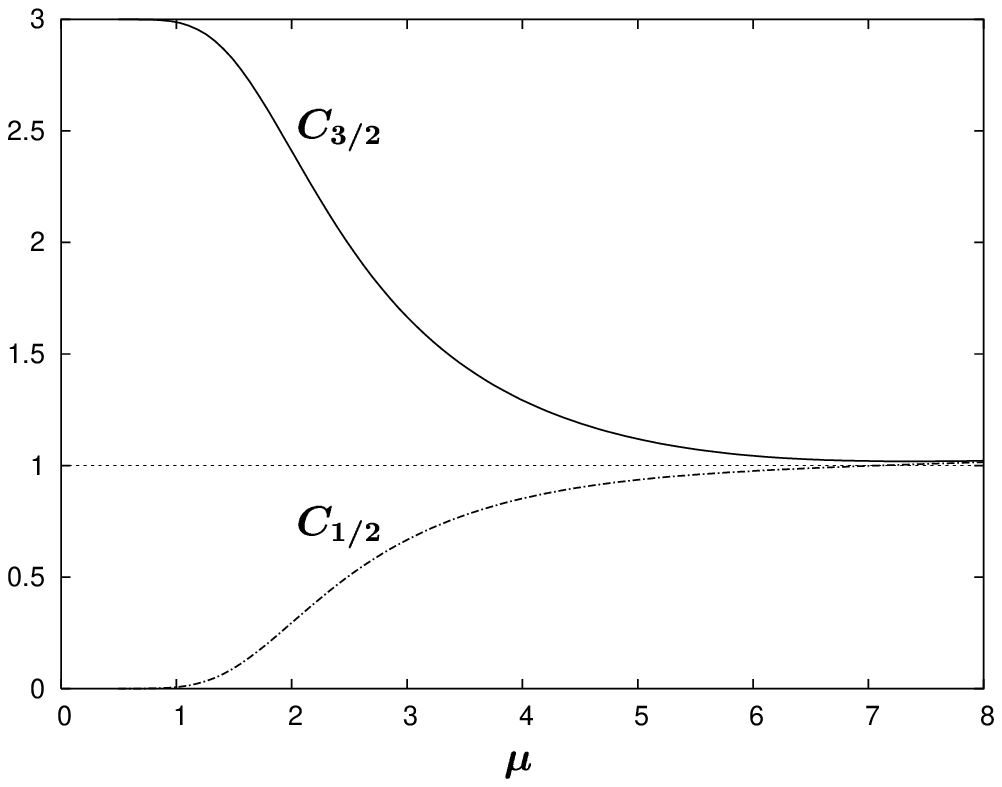}
\begin{center}
\mbox{Fig. 6. The plots of the asymptotic $C$-amplitudes
vs. $\mu$: $\gamma=3$.}
\end{center}
\end{figure}

\end{document}